\documentclass[aps,prl,twocolumn, titlepage,showpacs]{revtex4}

\usepackage{graphicx}
\usepackage{color}
\usepackage{dcolumn}

\usepackage{bm}

\begin{document}

\title{Superdiffusion of 2D Yukawa liquids due to a perpendicular magnetic field}

\author{Yan Feng}
\email{yanfengui@gmail.com}
\affiliation{Los Alamos National Laboratory, Mail Stop E526, Los Alamos, New Mexico 87545, USA}
\author{J. Goree}
\author{Bin Liu}
\affiliation{Department of Physics and Astronomy, The University of Iowa, Iowa City, Iowa 52242, USA}
\author{T. P. Intrator}
\author{M. S. Murillo}
\affiliation{Los Alamos National Laboratory, Los Alamos, New Mexico 87545, USA}

\date{\today}

\begin{abstract}

Stochastic transport of a two-dimensional (2D) dusty plasma liquid with a perpendicular magnetic field is studied. Superdiffusion is found to occur especially at higher magnetic fields with $\beta$ of order unity. Here, $\beta = \omega_c / \omega_{pd}$ is the ratio of the cyclotron and plasma frequencies for dust particles. The mean-square displacement ${\rm {MSD}} = 4 D_\alpha t^\alpha$ is found to have an exponent $\alpha > 1$, indicating superdiffusion, with $\alpha$ increasing monotonically to $1.1$ as $\beta$ increases to unity. The 2D Langevin molecular dynamics simulation used here also reveals that another indicator of random particle motion, the velocity autocorrelation function (VACF), has a dominant peak frequency $\omega_{peak}$ that empirically obeys $\omega_{peak}^2 = \omega_c^2+ \omega_{pd}^2/4$.

\end{abstract}

\pacs{52.27.Gr, 52.27.Lw, 66.10.C-}\narrowtext

\maketitle

\section{I.~Introduction}
Transport of charged particles under magnetic fields is important in studying plasma physics processes such as ion transport in tokamaks~\cite{Tsypin:1998} and the solar wind into Earth's magnetosphere~\cite{Hasegawa:2004}. An external magnetic field complicates the motion of all charged particles, as compared with the case without a magnetic field, so that their transport due to collisions is changed fundamentally. Kinetic theory including the effects of cyclotron motion~\cite{Spitzer:1956} is needed to study the collisional transport of plasmas with magnetic fields. Dusty plasmas provide an experimental and theoretical platform to study fundamental transport concepts.

Dusty plasma~\cite{Shukla:2002, Fortov:2005, Morfill:2009, Piel:2010, Bonitz:2010} is a four-component mixture of ions, electrons, gas atoms and electrically charged micron-sized dust particles. These dust particles are negatively charged, and their mutual repulsion is often described by the Yukawa or Debye-H\"uckel potential~\cite{Konopka:2000},
\begin{equation}\label{Yukawa}
{\phi(r) = Q^2 {\rm exp}(-r / \lambda_D) / 4 \pi \epsilon_0 r,}
\end{equation}
where $Q$ is the particle charge and $\lambda_D$ is the screening length due to electrons and ions. Due to their high particle charge, dust particles are strongly coupled, so that a collection of dust particles exhibits properties of liquids or solids. The size of dust particles allows directly imaging them and tracking their motion, so that collisional transport phenomena can be observed experimentally at the level of individual particles. Experiments can be performed either with a single horizontal layer of dust (2D) or with dust that fills a volume (3D). For the case of a 2D dusty plasma, which we will study, the electrons and ions fill a 3D volume, while the dust is constrained by strong dc electric fields to move only on a single plane~\cite{Feng:2011}.

Superdiffusion is a type of anomalous transport where particle displacements exhibit a scaling with time that is greater than for normal diffusion. When the time dependence of mean-square displacement of a particle is fit to the form
\begin{equation}\label{MSD}
{{\rm MSD}(t) = 4 D_{\alpha} t^{\alpha}}
\end{equation}
over times long enough for multiple collisions to occur, the signatures of normal diffusion and superdiffusion are $\alpha = 1$ and $\alpha > 1$, respectively. The coefficient $D_\alpha$ is not truly a diffusion coefficient if $\alpha > 1$, but nevertheless it is useful for quantifying the magnitude of random particle displacements.

For 2D systems, anomalous transport including superdiffusion has often been reported, for various {\it unmagnetized} systems including dusty plasmas. Indications of this kind of anomalous transport, attributed to low dimensionality, are often found in non-converging integrals for the random motion~\cite{Alder:1967, Alder:1970, Ernst:1970, Dorfman:1970, Donko:2009}. In a {\it magnetized} system, however, the trajectories of charged particles are fundamentally changed from those in an unmagnetized system, so it is an open question whether collisional particle motion is described as diffusion or superdiffusion. In this paper we seek to answer this question.

  Previous work has been reported for {\it unmagnetized} 2D dusty plasmas to assess whether $\alpha > 1$. This previous work includes experiments~\cite{Nunomura:2006, Liu:2008} and theoretical simulations~\cite{Liu:2007, Hou:2009, Ott:2009}.  Other transport coefficients that have been studied experimentally for 2D dusty plasmas include shear viscosity~\cite{Nosenko:2004, Feng:2011, Hartmann:2011} and thermal conductivity~\cite{Nunomura:2005, Nosenko:2008, Feng:2012a, Feng:2012b}. Simulations have also been reported for shear viscosity~\cite{Liu:2005, Donko:2006}, longitudinal viscosity~\cite{Feng:2013}, and thermal conductivity~\cite{Donko:2009, Hou:2009a, Kudelis:2013}.

Our paper is motivated by the recent attention given to dusty plasma behavior under magnetic field~\cite{Wang:2002, Filippov:2003, Jiang:2007, Dyachkov:2009, Banerjee:2010, Vasiliev:2011, Kahlert:2013, Kong:2013, Kopp:2014}. This attention is driven by experiments, which have only recently begun. There are at least three magnetized dusty plasma devices~\cite{Schwabe:2011, Reichstein:2012, Thomas:2012}, which now, or soon will be, producing experimental data. The prospects for these experiments has motivated simulations, including~\cite{Uchida:2004, Hou:2009b, Bonitz:2010b, Ott:2011a}, to study waves for 2D Yukawa liquids and solids under a magnetic field. In this literature, the magnetic field strength is quantified by a ratio of the cyclotron and plasma frequencies for dust particles, $\beta = \omega_c / \omega_{pd}$. For transport coefficients in magnetized strongly coupled plasmas, at the time we began writing this paper the literature included only studies for 3D systems, such as~\cite{Ott:2011} and one paper on a 2D Coulomb liquid~\cite{Ranganathan:2002}.

As we were finishing this paper, we learned of a new work, the first for diffusion in 2D Yukawa liquids, by Ott, L\"owen and Bonitz~\cite{Ott:2014}. Using a frictionless MD simulation, they determined the MSD for a wide range of time and $\beta$. Not claiming that it represented a diffusion coefficient, they reported a coefficient $D_\alpha$ evaluated at a particular time.  Our results complement those of~\cite{Ott:2014}. We investigate whether motion is superdiffusive, and we characterize a peak in the spectrum of the velocity autocorrelation function (VACF), which is another measure of random motion. Our simulation was not frictionless; we use a 2D Langevin MD simulation that includes the effects of gas friction, which are present in experiments.

We find that 2D motion of dust particles in a perpendicular magnetic field is superdiffusive when $\beta \approx 1$. We also find that the VACF has a spectrum that is dominated by a large peak due to a combination of  cyclotron motion and bouncing of particles within the cage defined by their neighbors. We find an empirical expression for the frequency of this peak.

\section{II.~Characterizing random motion}

We now review the measures of random motion that we use, the MSD (mean square displacement) and VACF (velocity autocorrelation function).

\subsection{A. MSD and superdiffusion}

Mean-squared displacement (MSD) characterizes self-diffusion~\cite{Einstein:1956}. It is defined as ${\rm MSD}(t) = \langle |\bf{r}_i ({\it t}) - \bf{r}_i ({\rm 0}) |^2 \rangle$, where $\bf{r}_i ({\it t})$ is the position of particle $i$ at time of $t$. Here, $\langle \rangle$ denotes the ensemble average over all particles and different initial times~\cite{Vaulina:2002}.

The MSD is a time series that reveals how random particle motion has different regimes, according to the time scale. For strongly coupled systems such as liquids, when the time is very short a particle moves mainly inside the cage formed by its nearest neighbors, which is called caged motion~\cite{Donko:2002}, the particle motion is termed ``ballistic''~\cite{Liu:2007}, and the MSD scales $\propto t^2$. At longer times, when several collisions have occurred, a particle can escape its cage and displace with a random walk described as self-diffusion. For these longer times, if the MSD time series is a straight line in a log-log plot, it is described by a power law, Eq.~(\ref{MSD}). In this equation, the factor of $4$ comes from the two dimensionality of our studied system, for 3D systems it would be $6$. Normal diffusion is characterized by $\alpha = 1$; while superdiffusion and subdiffusion are characterized by $\alpha > 1$ and $\alpha < 1$, respectively. Both superdiffusion and subdiffusion are also called anomalous diffusion.

A criterion is needed to judge whether motion is superdiffusive. Since data from simulations and experiments will never yield a value that is exactly $1$, some authors apply a more stringent criterion of $\alpha \ge 1.1$ for superdiffusion~\cite{Liu:2007}, instead of $\alpha > 1$. Another practical consideration is the time duration of the MSD data.  Indications have been reported~\cite{Ott:2009} that after a longer time interval, superdiffusive motion in a 2D Yukawa liquid vanishes, becoming  diffusive with $\alpha = 1$ at long times. Thus, it is desirable to assess the value of $\alpha$ for various time intervals, and to assess whether it trends to unity at long times, as we shall do in this paper.

\subsection{B. Velocity autocorrelation function}

Like the MSD, the VACF measures the temporal development of particles that are tracked individually, as they collide with others, but it is the fluctuating velocity rather than position that is used. The VACF is defined~\cite{Schmidt:1997} as the time series $\langle \bf{v}_i ({\it t}) \cdot \bf{v}_i ({\rm 0}) \rangle$, where $\langle \rangle$ also denotes the ensemble average over all particles and different initial times. If there are no magnetic fields, so that the particles move only because of their own inertia and collisions, the VACF will exhibit a damped oscillation, for strongly coupled systems such as a solid or liquid. The oscillation reflects the caging motion~\cite{Donko:2002} of a particle due to the interaction with its nearest neighbors.  In strongly coupled systems, diffusive motion arises from the gradual escape of a particle from a cage (the so-called  decaging motion) so that a particle becomes displaced. For normal diffusion, the VACF can be used to calculate the diffusion coefficient~\cite{Hansen:1986, Vaulina:2008, Dzhumagulova:2012}. We will use the VACF for another purpose because we will find that the motion is superdiffusive. In particular, we will use it to compute the vibrational density of states, which is the modulus of the Fourier transform of the VACF time series, plotted as a function of $\omega$~\cite{Schmidt:1997, Goncalves:1992, Teweldeberhan:2010}. This vibration density of states will reveal any preferred frequency for particle motion.

In this paper we add a magnetic field, and we expect that the time series for the VACF will oscillate, due not only to random interparticle interactions but also to cyclotron motion of individual particles. We expect that both kinds of oscillatory motion will be revealed in the vibrational density of states.

\section{III.~Simulation method}

We performed Langevin MD simulations, with additional Lorentz forces acting on dust particles due to the external perpendicular magnetic field. For each particle $i$, we integrate the Langevin equation
\begin{equation}\label{IntegratEq}
{m \ddot{\bf r}_i= Q \dot{\bf r}_i \times {\bf B} - \nabla \Sigma \phi_{i,j} - \nu m \dot{\bf r}_i + \zeta_i(t),}
\end{equation}
with a Lorentz force $Q \dot{\bf r}_i \times {\bf B}$, frictional drag ~\cite{Klumov:2009} $- \nu m \dot{\bf r}_i$ and a random force $\zeta_i(t)$. The random force $\zeta_i(t)$ is assumed to have a Gaussian distribution with a zero mean, according to the fluctuation-dissipation theorem~\cite{Feng:2008, Gunsteren:1982}. For the binary interaction potential $\phi_{i,j}$ we use the Yukawa repulsion, Eq.~(\ref{Yukawa}). Note that when there is a strong magnetic field, the dynamics of electrons and ions that account for the shielding may be completely changed~\cite{Schwabe:2011}, so that the interparticle interaction of 2D dusty plasmas may be more complicated. In this paper, we assume that the interparticle interaction is still the Yukawa interaction, as the zeroth order approximation.

We consider a uniform magnetic field in the $z$ direction perpendicular to the {\it x}-{\it y} plane in which the particles are constrained to move. We use the Langevin integrator of Gunsteren and Berendsen~\cite{Gunsteren:1982}. Time scales are normalized by the nominal plasma frequency, $\omega_{pd} = (Q^2/2 \pi \epsilon_0 m a^3)^{1/2}$~\cite{Kalman:2004}, which is also a time scale for interparticle collisions, in a system that is strongly coupled. Here, $m$ is the particle mass and $a \equiv (n \pi)^{-1/2}$ is the Wigner-Seitz radius~\cite{Kalman:2004} for an areal number density $n$. The magnetic field is characterized using the dimensionless parameter of $\beta = \omega_c / \omega_{pd}$, where $\omega_c$ is the cyclotron frequency of the dust particle. The time scale for gas frictional damping~\cite{Liu:2003} is chosen as $\nu = 0.027 \omega_{pd}$ to mimic typical experimental conditions~\cite{Feng:2008}, while the time scale for cyclotron motion is variable, by choosing $\beta$. We vary $\beta$ from $0$ to $1$, where the upper end of this range corresponds to an extremely strong magnetic field~\cite{Bonitz:2013}. For example, for a typical 2D dusty plasma experiment of~\cite{Feng:2010, Feng:2011} with $8$~micron diameter particles, $\beta = 1$ corresponds to a magnetic field of $B = 1.3 \times 10^4~{\rm T}$. Note that under stronger magnetic fields, plasma sources relying on capacitively coupled radio-frequency power can have some inhomogeneities. For example, filaments or enhanced ionization that are aligned parallel to the magnetic field lines were observed in~\cite{Schwabe:2011}. This nonuniformity of the plasma was observed to affect microparticle motion by causing an inhomogeneous ``pattern formation’’~\cite{Schwabe:2011}. We assume a spatially uniform plasma in our simulations, so that comparing our results to experiment must await a future experiment with conditions that are more uniform than in ~\cite{Schwabe:2011}. It is reasonable to anticipate such results because there are new facilities coming online that have the flexibility to alter the operation and design of their plasma source. We integrate Eq.~(\ref{IntegratEq}) using a time step of $0.037 \omega_{pd}^{-1}$, which we checked to be small enough for both the collisional and cyclotron motion.

The simulation parameters are chosen so that the collection of dust particles will behave as a liquid, according to the phase diagram of~\cite{Hartmann:2005}. To describe the dust particle charge, kinetic temperature $T$ and areal number density, we use the dimensionless quantities $\Gamma = Q^2/(4 \pi \epsilon_0 a k_B T)$ and $\kappa \equiv a / \lambda_D$. We choose $\Gamma = 200$ and $\kappa = 2$ as typical liquid conditions that are experimentally attainable using dusty plasmas.

We emphasize that for these parameters, in the absence of a magnetic field, it has been shown~\cite{Ott:2009} that motion is nearly that of normal diffusion, with $\alpha \approx 1$. We will determine how this conclusion changes as a magnetic field is added.

Another dimensionless parameter for magnetized dusty plasmas is the inverse Hall parameter for the dust $R_c = \omega_c / \nu$. When this ratio is much greater than unity, dust particles can complete circular orbits before the trajectory is disturbed by collisions with neutral gas, which occur at a rate $\nu$~\cite{Thomas:2012}. For the gas conditions simulated in our Langevin equation, $R_c = \beta / 0.027$.

Our simulation size is $N =1024$ particles constrained to planar motion in a rectangle of dimensions $65.5 a \times 56.7 a$. As in~\cite{Hou:2009, Ott:2011}, we use periodic boundary conditions. We truncate the Yukawa potential at radii beyond $22.9 a$ with a switching function to give a smooth cutoff between $22.9 a$ and $24.8 a$ to avoid an unphysical sudden force change when a particle moves a small distance~\cite{Feng:2013}. All simulation runs start from a random configuration of 1024 particles, then run $10^{5}$ steps to reach the steady conditions before starting recording data. After that, particle trajectories of the next $10^{7}$ steps are saved for data analysis. Note that the total time duration of $10^7 \times 0.037 \omega_{pd}^{-1} = 3.7 \times 10^5 \omega_{pd}^{-1}$ corresponds to $\approx 3.4$ hours for a typical value of $\omega_{pd} = 30~{\rm s}^{-1}$ in 2D dusty plasma experiments~\cite{Feng:2010, Feng:2011}, much longer than experimental runs. Representative trajectories are shown in Fig.~1. We verified that our simulations are free of any nonuniformity such as a flow or a localized peak in number density or kinetic temperature.

\section{IV.~Results}

\subsection{A.~Superdiffusion}

The calculated MSD time series for different $\beta$ values are presented in Fig.~2. As expected, displacements are reduced with an increasing magnetic field, i.e., an increasing $\beta$. After the initial ballistic portion, the MSD time series has its diffusive portion at longer times. For fitting the MSD data to determine $\alpha$, we will use the range of $100 < \omega_{pd} t < 1000$, which is for times later than the ballistic portion. We present the MSD curves two ways, normalized by the plasma frequency $\omega_{pd}$ and $f_c$ (where $f_c = \beta \omega_{pd}/2 \pi$ is the cyclotron frequency) in Fig.~2(a) and (b), respectively. In the latter we see that the oscillations in the MSD occur at the cyclotron frequency and its harmonics, indicating the dominant role of cyclotron motion at that time scale.

As our first main result, we present the exponent $\alpha$ in Fig.~3(a). For these $\Gamma$ and $\kappa$ conditions, we find that a 2D Yukawa liquid exhibits superdiffusion $\alpha = 1.1$ for a large magnetic field $\beta = 1$, and weak superdiffusion $1 < \alpha < 1.1$ for weaker magnetic fields. Without a magnetic field, $\beta = 0$, we find nearly normal diffusion, $\alpha \approx 1$, as was reported for previous unmagnetized simulations~\cite{Ott:2009}. Figure~3(a) shows that there is a monotonic trend for $\alpha$ to increase with $\beta$, i.e., for superdiffusion to become stronger as the magnetic field is increased. This result is different from the claim of Ranganathan {\it et al.} who simulated a 2D Coulomb liquid and reported normal diffusion~\cite{Ranganathan:2002}.

We find that the fitting exponent $\alpha$ depends slightly on the time range chosen for fitting. We chose four different fitting time ranges to detect how sensitive of the fitting exponent $\alpha$ is related to the time. From Fig.~3(a), as the time range for the fitting is longer, a clear trend that the exponent $\alpha$ is smaller can be easily detected. In a previous study of superdiffusion in 2D Yukawa liquids~\cite{Ott:2009}, Ott and Bonitz also found that the exponent $\alpha$ changes as they chose different time ranges to fit.

As we noted in the Introduction, the trajectories of random particle motion in a liquid are completely different in the presence of a magnetic field, so that before we conducted our simulations, there was no particular reason to expect motion to be either normal diffusion or superdiffusion. Our results in Fig.~3 make it clear that adding a magnetic field does cause superdiffusion.  The motion is nearly normal diffusion in the absence of magnetic field, but then it becomes weakly superdiffusive as a magnetic field is added with a small value of $\beta$, with the superdiffusion becoming stronger and reaching $\alpha = 1.1$ at a high magnetic field of $\beta = 1$. This superdiffusive tendency is not as powerful as in some cases, such as the unmagnetized simulation of \cite{Liu:2007} where $\alpha = 1.3$ was reported. While the superdiffusive tendency found here is less profound, there is no doubt that it is present for the time intervals that we studied: our results in Fig.~3 show very little scatter, and our fitting of the MSD curves in Fig.~2 that yielded the data in Fig.~3 had an exceptionally high coefficient of determination. Thus, we are confident in our empirical finding that motion is superdiffusive when magnetic field is added to a 2D strongly coupled plasma, when modeled as a Yukawa liquid in the presence of gas collisions, as in our simulation. We offer some discussion of this empirical finding in Sec.~IV~C.

We also characterize the coefficient $D_{\alpha}$ in Fig.~3(b). As in~\cite{Ott:2014}, we can see that $D_{\alpha}$ decreases monotonically as $\beta$ increases, meaning that the perpendicular magnetic field suppresses the self-diffusion of particles in 2D Yukawa liquids. We fit the data for $D_\alpha$ vs. $\beta$ for the fitting time range of $100 < \omega_{pd} t < 1000$ in Fig.~3(b) to an expression
\begin{equation}\label{Dfit1}
{D_\alpha = D_{\alpha 0} / (1+ \xi \beta)^2.}
\end{equation}
We chose the form of Eq.~(\ref{Dfit1}) so that it has an asymptotic behavior that is a constant $D_{\alpha 0}$ in the absence of magnetic field $\beta = 0$ and diminishes with the same scaling as classical diffusion $\propto 1/\beta^2$ for large magnetic field. For the range of $\beta$ that we explored, this expression fits the data well, with empirical coefficients $D_{\alpha 0} = 0.00616 a^2 \omega_{pd}$ and $\xi = 1.083$. This expression fits our data somewhat better than the expression derived from the Langevin equation by Ranganathan {\it et al.}~\cite{Ranganathan:2002}
\begin{equation}\label{Dfit2}
{D_\alpha = D_{\alpha 0} / (1+ \xi \beta^2).}
\end{equation}
Fits to both expressions are shown in Fig.~3. We note that these expressions, Eqs.~(\ref{Dfit1}) and (\ref{Dfit2}), each have two free parameters ($D_{\alpha 0}$ and $\xi$) and a tendency toward classical diffusion, which is different from the three-parameter fit used in~\cite{Ott:2014} which tends toward Bohm diffusion, $D_\alpha \propto 1/\beta$ for strong magnetic fields. We did not extend our simulation to large enough $\beta$ to test whether classical or Bohm diffusion better describes the transport because experiments might not be feasible at such a high magnetic field.

\subsection{B.~VACF peak frequency}

We can seek insight into the peculiarities of thermal motion under the partially magnetized conditions where we observed superdiffusion. To do this, we examine the velocity autocorrelation function (VACF), which is closely related to diffusion; its integral diverges for superdiffusion but converges for diffusion. Transforming the VACF in Fig.~4 to yield its spectrum, Fig.~5(a), our attention is drawn to the most prominent feature: a large peak. This peak is such a dominant feature of the VACF spectrum that it seems likely that to gain an understanding of the thermal motion, in the presence of both collisions and magnetic field, will require an understanding of the peak and its tendencies as the magnetic field is changed. Therefore we wish to characterize the frequency of the peak and its dependence on the parameters that characterize collisions ($\omega_{pd}$) and cyclotron motion ($\omega_c$).

In Fig.~4, we see that oscillations occur with or without magnetic field, but they are larger in amplitude and more persistent in time when the magnetic field is large. When there is a magnetic field, the oscillation frequency is close to the cyclotron frequency, as seen in Fig.~4(b) where time is normalized by $f_c^{-1}$. The decay of VACF is slower for stronger magnetic field, which is natural result of stronger cyclotron motion. Inspecting both Fig.~1 and Fig.~2, we also notice that, within a specific time range, the typical displacement of a particle under a stronger magnetic field is smaller. It seems that, under a stronger magnetic field, a particle needs a longer time to escape the cage formed by its nearest neighbors, i.e., a longer decaging time, due to stronger cyclotron motion.

The vibrational density of states~\cite{Goncalves:1992, Teweldeberhan:2010} is presented in Fig.~5(a). We calculated this as the spectral power of the VACF by a Fourier transformation of the normalized VACF time series. This vibrational density of states describes the collective motion of the particles. In Fig.~5(a) we see that this spectral power is not flat, but has a dominant peak of finite width. The prominence of this peak indicates that the thermal motion has a favored frequency.

As our second main result, in Fig.~5(b) we find that the peak frequency $\omega_{peak}$ increases with magnetic field $\beta$ according an empirical fit
\begin{equation}\label{peakf1}
{\omega_{peak}^2/ \omega_{pd}^2 = 0.25 + \beta^2,}
\end{equation}
or equivalently
\begin{equation}\label{peakf2}
{\omega_{peak}^2 = 0.25 \omega_{pd}^2 + \omega_{c}^2.}
\end{equation}
This expression combines two kinds of motion, collective motion at $\omega_{pd}$ and cyclotron motion of a single particle at $\omega_c$. Figure~5(b) and the expression Eq.~(\ref{peakf2}) illustrate how these two kinds of motion combine, for thermal motion of a strongly coupled plasma under magnetic field. There is a favored frequency, which is somewhat larger than the cyclotron frequency. We note that the coefficient of $0.25$ in Eq.~(\ref{peakf2}) was determined for our conditions, $\Gamma = 200$  and $\kappa = 2$; we have not determined whether it varies with those parameters. We can also express this peak frequency using the Einstein frequency $\omega_E$, which is the oscillation frequency that a charged particle’s motion would have in a cage formed by all the other particles, if all the other particles were stationary. From Fig.~2(b) in~\cite{Kalman:2004}, the Einstein frequency in our simulation conditions is $\omega_E \approx 0.35 \omega_{pd}$, so that we also find that this peak frequency can also be expressed as $\omega_{peak}^2 = 2 \omega_{E}^2 + \omega_{c}^2$, which is the same as the expression  for $\omega_{1,\infty}$  in~\cite{Bonitz:2010b}. For comparison, in Fig.~5(b) we also show the peak frequencies obtained from the frictionless simulation of~\cite{Bonitz:2010b}.

We note that our vibrational density of states is not the only way to characterize the frequency content of thermal motion. Another way, which has been widely used in the literature for strongly coupled plasmas, is the wave spectrum. For a 2D Yukawa liquid under a magnetic field, it has been used for example by Hou {\it et al.}~\cite{Hou:2009b}. To compare how our vibrational density of states and the wave spectrum quantify the frequency content, we present the wave spectrum in Fig.~6. We computed it using Eqs.~(2-4) of~\cite{Hou:2009b}, which use as their inputs the positions as well as velocities of particles, not just the velocities as in the vibrational density of states. Other differences are that the wave spectrum is resolved in both the magnitude and direction of the wave vector $k$. The direction refers to the particle velocity, as compared to the arbitrarily chosen direction of $k$, and it is said to be longitudinal or transverse according to whether $v$ is parallel or perpendicular to $k$, respectively. Examining the power spectra in Fig.~6, we see that the frequency content favors a peak frequency, which depends on the strength of the magnetic field, and the spectrum has a finite width about this peak frequency. The peak frequency is generally slightly higher than the cyclotron frequency, except when the magnetic field is absent, as was the case for our vibrational density of states. The wave number dependence, which is measured only by the wave spectra and not the vibrational density of states, shows how the wave becomes optical (i.e., $\omega$ does not approach zero for zero wave number) when a magnetic field is present. Note that, our obtained phonon spectra agree well with the experimental and simulation results in~\cite{Hartmann:2013}. As is well known for strongly coupled plasmas, the wave spectrum also shows how the wave starts as a forward wave, $d \omega / d k > 0$, for small $k$, but can become backward, $d \omega / dk < 0$, for larger $k$ where the wavelength is on the order of the particle spacing. Under a magnetic field, there is not only the dominant oscillation at a frequency somewhat above $\omega_c$, but also a lower-frequency mode at $\omega / \omega_{pd} \ll 0.5$. The latter mode was remarked upon by several authors for both 2D and 3D liquids under a magnetic field. Our Fig.~6 shows how this low frequency mode occurs more strongly for a transverse polarization.

\subsection{C.~Conceptual discussion of diffusion}

Our empirical findings, superdiffusion in the presence of a magnetic field and a VACF spectral peak that varies with the magnetic field, both hint at the complexity of random particle motion. This complexity arises from a combination of two kinds of particle motion: caged motion in a liquid and cyclotron motion in a magnetic field. To gain an appreciation for this complexity, we review here some of the concepts for diffusion in various physical systems, starting with some of the simplest ones. This discussion will lead us to recognize there is no obvious intuitive reason to expect normal diffusion, given the complex nature of the random motion for a liquid with magnetized particle motion.

Diffusion is often described as a process of random displacements for a specified time interval. The diffusion coefficient is estimated by dividing the square of the typical step-size displacement by the typical time interval between steps. Superdiffusion can happen when there is an unusual abundance of large displacements. L\'evy-flight displacements (in certain physical systems)~\cite{Sokolov:2000} are extreme examples of these large displacements, and they result in severe superdiffusion. More subtle increases in the abundance of large displacements will lead to a less severe superdiffusion.

For an electron or ion in a magnetized weakly-coupled plasma, there is a kind of normal diffusion called ``classical diffusion.'' The intuitive estimate for the classical diffusion coefficient is traditionally obtained by estimating the step size as the cyclotron radius and the time interval as the inverse Coulomb collision frequency. (There are several kinds of Coulomb collision frequencies; the relevant one is for perpendicular momentum deflection).

For Brownian motion of an isolated dust particle in gas, there is again a ``step size'' displacement between collisions, and a typical time between collisions. For the Brownian motion, the step size is the mean free path between collisions with gas atoms, and that is the only length scale. There is also only one time scale for the Brownian motion: the collision frequency with gas atoms. Displacements for a given time interval have a Gaussian distribution, and the resulting motion is diffusive.

For an unmagnetized strongly-coupled plasma, there is again a single length scale: the spacing between particles. (This is so unless there are modes present, which might have a particular wavelength and add another length scale.) There are two time scales of note: the Einstein time for oscillations in a cage, and a decaging time for a particle, which depends on temperature and structure. The latter time scale would lead to the diffusion. In the 3D case random motion should be diffusive in the absence of hydrodynamic flows. In the 2D case, however, there can be superdiffusion, which has been attributed to long-time correlations arising possibly from hydrodynamic modes, according to earlier literature for transport in 2D systems~\cite{Ernst:1970}.

Adding a sufficiently strong magnetic field to the strongly coupled plasma, gyration will provide an additional time and length scale. The additional length scale means that sometimes a diffusive step size might correspond to the cyclotron radius (as for classical diffusion in a weakly coupled plasma), or sometimes it might correspond to the interparticle spacing (as for a strongly coupled plasma without magnetic field). Or the step size might be some mixture of the two. There is no longer the simplicity of a single mechanism for random motion. This complex mixing of collective motion at $\omega_{pd}$ (without magnetic field) and cyclotron motion at $\omega_c$ (due to magnetic field) can be seen in our result for the vibrational density of state, Fig.~5(b). More than one mechanism is at play, so that there is no compelling reason to expect  that the step size of a displacement will be that of normal diffusion. Thus, there is no definitive reason to expect that self-diffusion will occur with the displacements increasing with time exactly as was the case for only one mechanism. In other words, the complexity of the random motion means that there is no simple reason for us to anticipate intuitively whether motion will be diffusive or superdiffusive. This situation leads us to rely on numerical simulations to provide an empirical answer.

\section{V.~Summary}

In conclusion, we have performed Yukawa MD simulations to study the diffusion and superdiffusion of 2D liquid dusty plasmas under a uniform perpendicular magnetic field. We characterized the stochastic motion of using the mean-squared displacement MSD, velocity autocorrelation function VACF, vibrational density of states, and phonon spectra. It is expected that adding a magnetic field will reduce the displacements of charged particles as they undergo collisions, and this indeed occurs. However, we also find that adding the magnetic field also changes the scaling of those displacements with respect to time so that the MSD scales with a greater power of time and the motion becomes superdiffusive. The vibrational density of states has only one dominant peak for all simulated conditions, and this peak frequency can be expressed a function of $\omega_c$ and $\omega_{pd}$. These conclusions may be tested in future experiments with magnetized dusty plasmas.

We thank M. Bonitz and T. Ott for valuable discussions and providing the data in Fig.~5(b). Work at LANL was supported by the LANL Laboratory Directed Research and Development program and Department of Energy contract No. W-7405-ENG-36, and work at Iowa was supported by National Science Foundation Grant No. 1162645.

\begin{figure}[p]
\caption{\label{trajectories} (Color online). Particle trajectories in the {\it x}-{\it y} plane for different constant perpendicular magnetic field strengths, $\beta = \omega_c / \omega_{pd} = 0$ (a), 0.5 (b), and 1.0 (c). The random walk without a magnetic field changes its character, becoming more circular and wander less as the magnetic field increases in (b) and (c). Color represents time, and only $\approx 10\%$ of the simulated region and $\approx 0.03\%$ of the simulation duration are shown here. Our simulation conditions are $\Gamma = 200$ and $\kappa = 2$.}
\end{figure}

\begin{figure}[p]
\caption{\label{MSD} (Color online). Mean-squared displacement MSD for different magnetic fields, in the unit of $\omega_{pd} t$ (a), or $f_c t$ (b). At long times $100 < \omega_{pd} t < 1000$, well after the ballistic portion, we can fit the MSD time series to Eq.~(2). As $\beta$ increases, the MSD curves are lower and lower, indicating that the wandering motion of particles is suppressed by magnetic field, and $D_{\alpha}$ decreases. Oscillations at shorter times are due  to cyclotron motion of individual particles, as is best seen in (b), where time is normalized by $1/f_c$. Dips in the MSD time series occur around one cyclotron period, two periods, and so on. Here, the MSD is normalized using the Wigner-Seitz radius $a$. Our simulation conditions are $\Gamma = 200$ and $\kappa = 2$.}
\end{figure}

\begin{figure}[p]
\caption{\label{MSDfit} (Color online). (a) Indication of superdiffusion. The exponent $\alpha$ is $> 1$, especially for strong magnetic fields $\beta \approx 1$. These data are the results of fitting the MSD time series in Fig.~2 to the exponential scaling of Eq.~(\ref{MSD}) in the indicated time ranges. This result $\alpha > 1$ for a 2D Yukawa liquid is different from that of Ranganathan {\it et al.} who reported normal diffusion, $\alpha = 1$, for a 2D Coulomb liquid~\cite{Ranganathan:2002}. As the time range for the fitting is longer, we can see a clear trend that the exponent $\alpha$ is smaller. In (b), as $\beta$ increases from 0 to 1, the coefficient $D_{\alpha}$ decreases monotonically more than $70\%$ as $\beta$ increases, which means that the magnetic field greatly suppresses the wandering of particles. Note that the scatter of our data for each $\beta$ value corresponds to the error bar. Fits of our $D_{\alpha}$ data for the time range of $100 < \omega_{pd} t < 1000$ only to our empirical expressions Eq.~(\ref{Dfit1}) and Eq.~(\ref{Dfit2}) derived by Ranganathan {\it et al.}~\cite{Ranganathan:2002} are shown as solid and dashed lines, respectively.}
\end{figure}

\begin{figure}[p]
\caption{\label{VACF} (Color online). Velocity autocorrelation function, VACF, for different magnetic fields. Time is normalized by $1/\omega_{pd}$ in (a) and $1/f_c$ in (b). The oscillations decay more slowly with higher magnetic field, as seen in (a) for increasing $\beta$. These oscillations are mainly due to the cyclotron motion, since its frequency is nearly the same as $f_c$. The period of oscillation is related to the magnetic field strength, as seen in (b) where VACF curves for two values of $\beta$ are nearly aligned.}
\end{figure}

\begin{figure}[p]
\caption{\label{VACFspectra} (Color online). (a) Vibrational density of states, i.e., spectral power of the normalized VACF for different magnetic fields. The curves exhibit a dominant peak; the frequency of this peak is plotted in (b). The peak frequency increases monotonically as $\beta$ increases. The peak frequency fits an empirical curve, $\omega_{peak}^2/ \omega_{pd}^2 = 0.25 + \beta^2$, i.e., $\omega_{peak}^2 = 0.25 \omega_{pd}^2 + \omega_{c}^2$, shown as a smooth curve. This fit shows how the peak frequency is always greater than the cyclotron frequency. For comparison, we also plot the data from the frictionless simulation of~\cite{Bonitz:2010b} for the peak frequency of the longitudinal waves at the wave number of $ka = 5.55$.}
\end{figure}

\begin{figure}[p]
\caption{\label{phononspectra} (Color online). The longitudinal and transverse phonon spectra of our 2D Yukawa liquid when $\beta = 0$ (a,b), $\beta = 0.5$ (c,d), $\beta = 1$ (e,f). These spectra differ from the the vibrational density in Fig.~5 because they reflect both spatial and temporal fluctuations as characterized by a current~\cite{Hou:2009b}, not just the temporal fluctuations characterized by the VACF. }
\end{figure}

\end{document}